# Effect of crystallographic anisotropy on the resistance switching phenomenon in perovskites

T. Plecenik[a], M. Tomášek[b], M. Belogolovskii[c], M. Truchly[a], M. Gregor[a], J. Noskovič[a], M. Zahoran[a], T. Roch[a], I. Boylo[c], M. Španková[d], Š. Chromik[d], P. Kúš[a] and A. Plecenik[a]

[a]Department of Experimental Physics, Faculty of Mathematics, Physics and Informatics, Comenius University, 84248 Bratislava, Slovak Republic
[b]Institute of Physical Chemistry and Chemical Physics, Faculty of Chemical and Food Technology, Slovak University of Technology, 81237 Bratislava, Slovak Republic
[c]Donetsk Institute for Physics and Engineering, NASU, 83114 Donetsk, Ukraine
[d]Institute of Electrical Engineering, Slovak Academy of Sciences, 84104 Bratislava, Slovak Republic

Resistance switching effects in metal/perovskite contacts based on epitaxial c-axis oriented $YBa_2Cu_3O_{6+c}$ (YBCO) thin films with different crystallographic orientations have been studied. Three types of Ag/YBCO junctions with the contact restricted to (i) c-axis direction, (ii) ab-plane direction, and (iii) both were designed and fabricated, and their current-voltage characteristics have been measured. The type (i) junctions exhibited conventional bipolar resistance switching behavior, whereas in other two types the low-resistance state was unsteady and their resistance quickly relaxed to the initial high-resistance state. Physical mechanism based on the oxygen diffusion scenario, explaining such behavior, is discussed.

The resistive switching (RS) effect observed in capacitor-like metal/insulator/metal junctions belongs to the most promising candidates for the next generation of the memory cell technology based on a sudden change of the junction resistance caused by an electric field applied to the metal electrodes. Despite a burst of activities triggered by increasing technological interest for this phenomenon[1,2], details of its physical mechanism are still debated. The most studied are highly insulating binary transition metal oxides, where a key ingredient of the RS effect is believed to be a redistribution of oxygen ion vacancies which results in the formation (low-resistive state – LRS) and rupture (high-resistive state – HRS) of conductive filaments within the insulating media.[1,2] For more complex oxides like perovskites, the process of the RS effect seems to be more complicated. In this case the switching was usually bipolar whereas in simple binary oxides bi- and unipolar behavior has been observed.[3-5]

In the following we are studying the RS effect in contacts based on conducting $YBa_2Cu_3O_{6+c}$ (YBCO) films, compounds with a crystal structure which promotes electronic transport primarily within a two-dimensional layer of atoms. One of the ways to shed light on their basic properties is to study an electric-field effect in the normal state. Up to now, two mutually exclusive field-effect mechanisms have been proposed. The first one, fundamentally electronic, is based on a conventional approach which takes into account the Coulomb interaction of an applied electric field with mobile current carriers.[6] It is fast, symmetric with respect to the bias polarity and results in the enhancement or depletion of the number of charge carriers within a few near-surface atomic layers.[6] The second mechanism[7] is related to direct interaction of oxygen ions with an applied electric field which causes significant charge carrier rearrangement due to a comparatively small oxygen migration energy and high density of vacancies in the oxygen sub-lattice in an optimal doping state. Such a process is characterized by a slower time constant and can be unequal in magnitude at positive and negative voltage biases of identical absolute value.[7] To choose between the two mechanisms of field-induced changes in normal-state cuprates is an experimental task. Existing data cannot uniquely support any explanation since some evidences support direct doping of superconducting $CuO_2$ planes by electric-field induced mobile charges[8] whereas others favor the oxygen-diffusion scenario.[9]

In our study, we have employed a strong crystallographic anisotropy of the YBCO compound in order to find new arguments favoring one of the electric-field mechanisms. It should be noticed that although the RS effect was already observed in YBCO-contacts[3,10-14], up to our knowledge the influence of the crystallographic anisotropy has not yet been consistently studied. We have found a very distinct RS behavior in contacts restricted to (i) c-axis direction, (ii) ab-plane direction and (iii) both. In the latter two cases we observed a strong asymmetry of current-voltage (I-V) curves with respect to voltage V as was predicted in Ref. 7. We interpret the observations presented below as a result of the oxygen migration towards to cuprate/metal interface (LRS) or from the interface (HRS) depending on the electrodes bias polarity.

The c-axis oriented YBCO films were prepared on single-crystalline $LaAlO_3$ (001) substrates by high-pressure on-axis dc magnetron sputtering described elsewhere.[4] To characterize the influence of the crystallographic orientation of the YBCO film on the RS effect, three types of YBCO/Ag junctions were fabricated. The type (i) junctions with contact only in

---

Author to whom correspondence should be addressed. Electronic mail: tomas.plecenik@fmph.uniba.sk

the *c*-axis of the YBCO film were obtained as follows: The YBCO 100 μm wide strip electrodes were formed by optical lithography and subsequent Ar ion etching. The next step was to deposit a SiO$_2$ layer by vacuum evaporation and to remove a photoresist. After that, the lift-off photolithography was used to fabricate 10 μm wide Ag strips across the YBCO layers. Before thermal evaporation of the 110 nm thick Ag upper electrode, the YBCO surface was cleaned by additional ion beam etching during 10 min. The type (ii) junctions with contact only in the YBCO *ab*-plane were prepared in a similar way, but the SiO$_2$ layer was deposited before the 100 μm wide strip YBCO electrodes were patterned. Finally, the type (iii) junctions with contact in both *c*-axis and *ab*-plane were prepared under identical procedures but without the SiO$_2$ deposition step (see inset in Fig. 2).

Transport measurements have been done in a standard four-contact setup (see the inset in Fig. 1) with a Keithley 220 Programmable Current Source and Keithley 2000 Multimeter controlled by a computer via GPIB. The current-voltage characteristics were measured by applying periodical saw sweeping current with a maximal value of order of tens of μA and a time period of about 10-15 minutes (see the inset in Fig 1). Typical *I-V* curves of the three types of junctions are shown in Fig. 1. Initially, all three types of the junctions were in the HRS (~ 100 kΩ at 100 mV, branches 1 and 4 on Fig. 1). The *I-V* curves of the type (i) junctions with a contact only in the *c*-axis of the YBCO film exhibited classical bipolar RS behavior. After applying sufficiently high negative current/voltage to the YBCO film, the junction switched from the HRS (branch 1 on Fig. 1) to the LRS (~ 10 kΩ at 100 mV, branch 2) and remained in this state until sufficiently high positive current/voltage was applied (transition from LRS branch 3 to HRS branch 4 on Fig. 1). The other two types of junctions with a contact also (or only) within the *ab*-plane of the YBCO film were capable to switch to the LRS as well. But this state was unstable within the used time frame and due to fast oxygen diffusion within the *ab*-plane the distribution of oxygen vacancies gradually relaxed back to the initial HRS state once the absolute value of negative voltage biases started to decrease. At zero voltage the junctions had their highest possible resistances and thus no transition to a more resistant state was observed at positive biases. Moreover, a slight shift to a less resistant state visible in (ii) and (iii) junctions (compare two hysteretic curves in the first quadrant of Fig. 1) can originate from self-heating effects.

Classical scenario which describes the bipolar RS effect on the base of the oxygen-subsystem dynamics (type (i) junction in Fig. 1) assumes electrically-enhanced diffusion of oxygen ions within the contact area.[11,13,14] Initial state of the junctions is high resistive one due to the presence of higher amount of oxygen vacancies at the YBCO surface[15-17]. The greater is the amount of oxygen vacancies per unit of volume, the higher is the resistivity $\rho$.[18,19] Application of the electric field enhances the diffusion of negatively charged oxygen ions towards to interface or from the interface due to actual polarity. However, the migration of oxygen ions becomes considerable only above certain field intensity which accelerates oxygen species enough to overcome the activation energy $E_A$ of their motion to a next cell[20]. Observation of high-$T_c$ superconductivity suppression and restoration near the Ag/YBCO interface during the RS effect at temperatures below $T_c$, supports this scenario as the variation of oxygen content modifies electrical properties of the near-interface YBCO layer from insulating to superconducting.[14,17,21] Generally, both the resistivity $\rho$ and activation energy $E_A$ depend on the oxygen content as well as on the direction inside the YBCO crystal. Typical value of the $\rho_c/\rho_{ab}$ ratio is about 10 ÷ 100 depending on oxygen content where $\rho_{ab}$ is the resistivity in the in-plane (*ab*-plane) direction and $\rho_c$ in the out-of-plane (*c*-axis) direction[19].

Standard Arrhenius relation defines the diffusion coefficient $D$ as $D = D_0.\exp(-E_A/k_BT)$ where $D_0$ is a constant for a certain temperature range and crystal direction. Though exact dependences of $E_A$ and $D_0$ on the above mentioned parameters are not known, it has been found that the ratio $D_{ab}/D_c$ is usually much greater than $10^3$.[22-24] In our opinion, just the difference between in-plane and out-of-plane diffusion coefficients is in large extent responsible for the different RS behavior of type (i) samples and those of types (ii) and (iii). As the in-plane $D_{ab}$ is considerably higher, the oxygen species can easily re-distribute back to the equilibrium HRS state much faster after the bias voltage is decreased or turned off. For illustration, we consider such an evolution running according to one-dimensional diffusion equation (neglecting the electric field at low biases) with analytical solution c(x,t)=(const./$\sqrt{4\pi Dt}$ )exp(-x$^2$/4Dt), where *c(x,t)* stands for oxygen concentration at place *x* and time *t* (this equation was used to calculate diffusion profiles in Fig. 2). From this equation it follows that the time scale of a diffusion process may be estimated using simple relation $L^2 \approx Dt$, where $t$ is the time necessary to diffuse over the distance $L$. Assuming that the depth of the near-contact area $L$ where the RS effect takes place is similar for both directions (in-plane and out-of-plane), we directly obtain the ratio between two time scales being $t_c/t_{ab} = D_{ab}/D_c \gg 10^3$ (Fig. 2). Usually, the oxygen deficient layer at the Ag/YBCO interface where the RS takes place is few nanometers in size – let us take $L \sim 3$ nm.[17,20] According to the time scale of our experiment (Fig. 1), the time necessary to diffuse over distance $L$ for the faster in-plane diffusion process is several seconds or lower. If we take $t_{ab} \sim 10$ s, then for the in-plane diffusion coefficient we get $D_{ab} = L^2/t_{ab} \sim 10^{-14}$ cm$^2$·s$^{-1}$. Using the value of $t_c/t_{ab}$ ratio, the corresponding time for the out-of-plane diffusion should be at least $t_c \sim 10^4$ s, hence $D_c = L^2/t_c \sim 10^{-17}$ cm$^2$·s$^{-1}$. Such values are in good correspondence with experimentally obtained diffusion coefficients.[22,25] Thus, significantly lower diffusion coefficient $D_c$ in out-of-plane direction preserves metastable LRS state for considerably longer times, what was also confirmed experimentally by measuring the relaxation time in the LRS state (Fig. 2). However, it could be anytime switched back to the HRS by the enhanced electro-diffusion process for a critical electric field intensity of the opposite polarity.

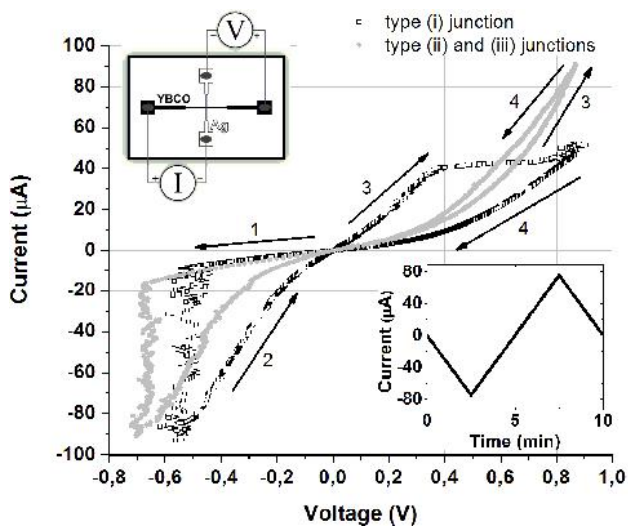

*Fig. 1: Typical I-V curves of the three types of YBCO/Ag junctions. Upper left inset: Schematic drawing of the measurement. Bottom right inset: Typical time profile of the applied current.*

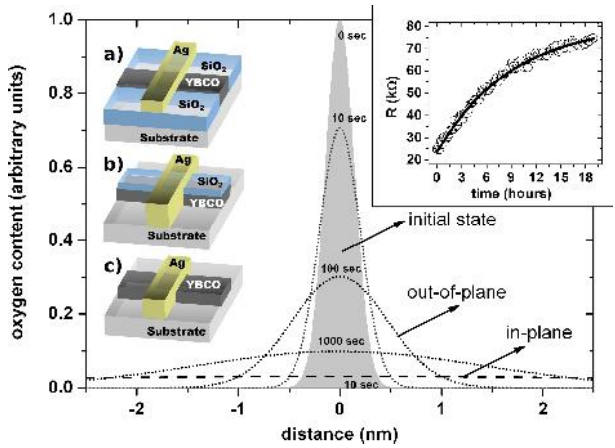

*Fig. 2: (Color online) Comparison of time scales for in-plane and out-of-plane diffusion processes. Initial LRS state represents an oxygen structure around 1 nm in size which is located at x = 0 (filled area) – this state is not stable and undergoes the diffusion process. Dotted lines visualize snapshots of such process using one-dimensional diffusion equation for c(x,t) (see text) after 10, 100, and 1000 seconds in the out-of-plane direction ($D_c$) and dashed line after 10 seconds in the in-plane direction ($D_{ab}$) from the same initial state. Left inset: Schematic drawing of the YBCO/Ag junctions with contact made: a) only in the c-axis direction – type (i), b) only in the ab-plane direction – type (ii) and c) in both directions – type (iii). Right inset: Typical relaxation curve from LRS back to equilibrium HRS for a junction with a contact only in the c-axis direction. The typical relaxation time constant estimated from the exponential decay fit was about 9 hours in qualitative agreement with our oxygen diffusion model. The measurements were done at 100 nA current and ~ 1-10 mV voltage biases insufficient to cause any resistance switching. The relaxation times may, however, vary from sample to sample depending on the oxygen vacancy concentration, interface quality and other factors.*

To conclude, an effect of the YBCO crystallographic anisotropy on the resistance switching effect has been studied in YBCO/Ag contacts with well controlled direction of the charge current in the contact area. By measuring current-voltage characteristics we have shown that the crystallographic orientation of the YBCO compound has a strong influence on the resistance switching behavior. Whereas a conventional bipolar resistance switching has been observed on c-axis oriented YBCO/Ag contacts, fast oxygen diffusion processes in the ab-plane are preventing the ab-plane junctions from persisting in the low-resistance state. Such junctions thus tend to relax to the initial high-resistance state within several seconds after the bias voltage is turned off and no stable resistance switching has been observed. The observations are interpreted in terms of the oxygen-diffusion scenario and provide, in our opinion, new arguments in favor of the mechanism of electric field effect in high-$T_c$ films which assumes direct interaction of the applied electric field with oxygen ions.

This work was supported by the Slovak Research and Development Agency under Contract No. LPP-0176-09 and is also the result of the project implementation: Grant No. 26220220004 supported by the Research & Development Operational Programme funded by the ERDF. One of us (M.B.) is grateful to Comenius University for hospitality during the course of this work.